\documentclass[mathleft
]{an}
\usepackage{graphicx}
\usepackage{times}
\begin{document}

\Pagespan{789}{}
\Yearpublication{2006}%
\Yearsubmission{2005}%
\Month{11}%
\Volume{999}%
\Issue{88}%

\title{APOGEE: The Apache Point Observatory Galactic Evolution Experiment}

\author{C. Allende Prieto\inst{1}\fnmsep\thanks{Corresponding author:
  \email{cap@mssl.ucl.ac.uk}\newline}
S.~R. Majewski\inst{2},
R. Schiavon\inst{3},
K. Cunha\inst{4},
P. Frinchaboy\inst{5,6},
J. Holtzman\inst{7},
K. Johnston\inst{8},
M. Shetrone\inst{9},
M. Skrutskie\inst{2},
V. Smith\inst{4} 
\and J. Wilson\inst{2}
}
\titlerunning{APOGEE}
\authorrunning{Allende Prieto et al.}
\institute{
Mullard Space Science Laboratory, University College London, 
Holmbury St. Mary,  Surrey RH5 6NT, UK
\and 
Department of Astronomy, University of Virginia,
P.O. Box 400325, Charlottesville, VA 22904-4325, USA
\and 
Gemini Observatory, 670 North A`ohoku Place, Hilo, HI 96720, USA
\and
National Optical Astronomical Observatory, Casilla 603, La Serena,  Chile
\and
NSF AAP Fellow, Univ. of Wisconsin-Madison, Department of Astronomy, 4506 Sterling Hall,
 475 N. Charter Street, Madison, WI 53706, USA
\and
Any opinions, findings, and conclusions or recommendations 
expressed in this material are those of the author and 
do not necessarily reflect the views of the National Science 
Foundation (NSF).
\and
New Mexico State University, Las Cruces, NM 88003, USA
\and
Astronomy Department, Columbia University, New York, NY 10027, USA
\and
McDonald Observatory, University of Texas, Austin TX 78712, USA
}

\received{}
\accepted{}
\publonline{later}


\abstract{
APOGEE is a large-scale, NIR, high-resolution ($R\sim$ 20,000) spectroscopic survey of Galactic stars. It is one of the four experiments in SDSS-III. 
Because APOGEE will observe in the $H$ band, where the extinction is six times smaller than in $V$, it will be the first survey to
pierce through Galactic dust and provide a vast, uniform database of chemical abundances and radial velocities for stars 
across all Galactic populations (bulge, disk, and halo). The survey will be conducted with a dedicated, 300-fiber, cryogenic, spectrograph that is being built at the University of Virginia, coupled to the ARC 2.5m telescope at Apache Point Observatory. APOGEE will use a significant fraction of the SDSS-III bright time
during a three-year period to observe, at high 
signal-to-noise ratio ($S/N>$100), 
about 100,000 giant stars selected directly from 2MASS down to a 
typical flux limit of $H<13$. 
The main scientific objectives of APOGEE are: (1) measuring unbiased metallicity distributions and abundance 
patterns
for the different Galactic stellar populations, (2) studying the processes of star formation, feedback, and chemical mixing in the Milky Way, (3) surveying the dynamics of the bulge and disk, placing constraints on the nature and influence of the Galactic bar and spiral arms, and (4) using extensive chemodynamical data, particularly in the inner Galaxy, to unravel its formation and evolution.
}

\maketitle

\section{Introduction}

Our knowledge of the structure, formation and evolution of galaxies is built upon observations of numerous galaxies 
as well as on detailed measurements in the Milky Way.  Only for the Galaxy can we quantify precisely the basic properties
of individual stars from parallaxes, proper motions and high-resolution spectroscopy.
Only for the Milky Way we can make a census of minority stellar populations which,
as in the case of the stellar halo, can be central to understanding the earliest
epochs of galactic formation. In addition, the spatial resolution of gas and
dust maps available for the Galaxy is orders of magnitude better than for
extragalactic systems. 

As a result of developments in instrumentation over the last decades, 
high-resolution spectroscopic studies have become more and more precise, 
involve larger samples, and reach fainter magnitudes. These advances have made 
it possible to carry out a thorough study of the stellar populations 
in the solar neighborhood and measure high precision chemical abundances for dozens of elements and hundreds 
of stars. These observations have revealed, among other things, a clear dichotomy
between the chemical patterns in thin- and thick-disk stars, despite their 
overlapping distributions in phase space. 

Progress has probably been even more spectacular for studies using lower spectral
resolution, thanks to massive multiplexing on fiber spectrographs such as
those used in the Sloan 
Digital Sky Survey (SDSS; York et al. 2000), which operate at $R=$ 2,000. 
The information per spectrum is
more limited, but this is compensated by deeper and much larger samples of stars.  More than
half a million stars have already been observed in the course of the SDSS,
and this number will double over the next three years. Several tens of thousands 
of stellar spectra have already been analyzed in recent studies from SDSS 
(Allende Prieto et al. 2006, Ivezi\'c et al. 2008), 
and higher resolution ($R\sim$ 7,500) spectra, over a limited spectral
window, are now becoming available from the Radial Velocity Experiment (RAVE;
Zwitter et al. 2008).
This allows a thorough characterization of the properties of  
the Milky Way stellar halo, and provide information on the thick disk 
over a wide range of galactocentric distances. Now we know that the thick-disk
stars are old, but intermediate in age between the halo and the thin disk, 
and that the thick disk does not show radial abundance gradients such as those well-known
in the thin disk  (Allende Prieto et al. 2006).

In stark contrast with the fast pace at which we have been able to make
progress in our exploration of the outer Galaxy at low resolution
and the steady accumulation of high resolution information on nearby stars,  
the situation for understanding the non-local, inner Milky Way at any resolution has been much less favorable.
The high extinction in the Galactic mid-plane and
towards the Galactic center severely complicates observational studies 
at traditional optical wavelengths. Spectroscopic
studies of the Galactic bulge have been limited to small numbers of stars 
--- almost exclusively in Baade's window.
The Apache Point Galactic Evolution Experiment (APOGEE) is targeted to dramatically
change this situation by using highly-multiplexed near-IR 
high-resolution spectroscopy
to pierce through dust and probe stars in the central parts of the galaxy at high resolution.  In addition,
APOGEE will push the systematic, high resolution, high throughput 
spectroscopic study of Galactic stars
into the outer Galaxy, creating a uniform spectroscopic database for stars in all 
populations of the Milky Way.  Such homogeneous but extensive
datasets are necessary to produce an integrated 
picture of the chemical and kinematical evolution of our home galaxy.

\section{Why in the H band? Why now?}

We have mentioned that the most serious problem to access the
central parts of the Galaxy is dust obscuration.  Dust absorption in the
$H$ band ($\sim 1.6 ~\mu$m) is more than 5 times 
smaller than in the optical (e.g., $A_H/A_V=0.16$).
A second key fact is that near-IR instrumentation has now reached the level 
of maturity necessary to make wide-field, multiplexed 
high-resolution spectroscopy at these wavelengths both feasible,
and very efficient. In just a year of observing time (bright time during three years) 
APOGEE will make a substantial leap
in the total number of available, high resolution
optical or NIR spectra for Galactic stars in general
(about two orders of magnitude!).

In addition, several other factors make this an ideal time to undertake an $H$-band 
survey of giant stars in the Galaxy with the Sloan telescope:

\begin{itemize}
\item K-type giant stars down to the magnitude of the red clump are bright enough in the $H$ band 
	that the sensitivity of current IR detectors and throughput of NIR gratings allow 
	us to reach the central parts of the Milky Way at $R\sim$ 20,000 and $S/N=$100
	with a mid-class telescope (e.g., 2.5-m aperture).

\item	A low atmospheric extinction in the infrared makes it possible to access
	the Galactic bulge from the Northern Hemisphere (Apache Point, in particular).
	
\item A complete catalog of sources down to $H\sim15$ is available to carry 
out  
an unbiased target selection (2MASS; Skrutskie et al. 2006).
	
\item The software/hardware infrastructure deployed for the SDSS and the 
	experience gained in large survey operations provide a foundation on which
	efficiently to build  
	a massive spectroscopic
	data set and make it accessible to the scientific community. 
	
\item The schedule for the European mission Gaia plans a release of its 
	comprehensive astrometric catalog by 2017 (parallaxes and proper motions for
	1 billion stars with an accuracy of $\sim 10-25 \mu$ arcsec!)\footnote{Note
	that Gaia will fly a high-resolution spectrograph. 
	However, it will operate at shorter wavelengths (847-874 nm) 
	and, with an integration 
	time limited by the continuous rotation of the satellite, 
	chemical abundances will only be measured down to $V<12-13$ (Wilkinson et al.
	2005), equivalent to $H<9$ for a late-type giant.}.  Thus, we can assemble complete
	kinematical data and accurate distances for stars for which APOGEE will derive 
	precision chemistry and radial velocities.

\end{itemize}

In addition, the $H$-band accesses a part of the spectral energy distribution where
the continuum is form deepest and which contains lines formed close to LTE 
conditions for multiple interesting elements, including C, N, O, Fe, several  
$\alpha$ elements, iron peak elements, and odd-$Z$ elements. 

\section{Scientific objectives}

High resolution spectroscopy allows the characterization of
 a resolved stellar population
in exquisite detail. The chemical compositions of FG-type stars  can be routinely derived from optical spectra 
for many elements with a precision better than 0.04 dex 
(see, e.g. Bensby et al. 2007; Fuhrmann 2008; Reddy et al. 2007). 
In the NIR, the presence of
significant telluric absorption, and emission features, mainly OH, from high
atmospheric layers, complicates matters, but
previous studies in
this band have successfully
determined chemical abundances for an array of elements with a relative
uncertainty of $<0.1$ dex (e.g. Cunha \& Smith 2006, Rich \& Origlia 2006,
 Mel\'endez et al. 2008). Given that APOGEE will have considerable
advantages regarding calibration being such a large and homogeneous survey, we
expect to be able to reach this mark easily. 

By obtaining, with this kind of precision,  
the chemical abundances for the elements C, N, O, Mg, Al, Si, Ca, Ti, Cr, Fe
and Ni for all surveyed stars, as well as Na, S, V, Mn, Co and other elements for brighter stars,  APOGEE will be able to dissect the chemical history of 
the different Galactic stellar populations.
APOGEE will not only map the metallicity
distribution functions and differential abundance patterns
for these elements as a function of Galactic position, 
but measure their relative rates of enrichment, correlations with kinematics, and the chemical/kinematical
relationships between populations.

A major objective of APOGEE is to penetrate the bulge of the Milky Way, uncover
the dynamics of the Galactic bar, and find out its connection to the dynamics
of the disk. The presence of a bar in the Milky Way has been inferred 
from photometry (see L\'opez-Corredoira et al. 2007 and references therein). 
Radial velocities, which should be free of systematic errors at 
a level $<1$ km s$^{-1}$, can be extracted by cross-correlation from the APOGEE
spectra, and  should confirm the presence of the bar, as well as provide,
by comparison with dynamical models, fair estimates of its mass and hints
about its origin.

APOGEE will also target areas of the Milky Way other than the bulge. 
Although the local thin
disk, thick disk, and halo have been studied before, 
APOGEE will reach 
other parts of the disk far from the solar circle, and obtain a larger
and homogeneous sample of high resolution spectra of distant halo giants. Most importantly,
APOGEE will, for the first time, place the measurements for all the 
stellar populations in the Galaxy under the same system, removing systematic
effects that plague existing studies based on compiling observations
from different instruments and analysis protocols.

Another goal of the survey will be to measure the rotation curve
of the Milky Way.  Using stars will make it unnecessary to base the distance
estimates on the assumption of circular motion as it is done for 
observations of gas at longer wavelengths, for we can rely on the
spectroscopic parallaxes derived directly from measured surface gravities, metallicities and temperatures.
In addition, from these well-established spectroscopic parallaxes we can
explore the 3-D distribution of dust, expressed in photometric colors.

\section{Instrument design}

\begin{figure}
\includegraphics[width=50mm,angle=90,height=5.7cm]{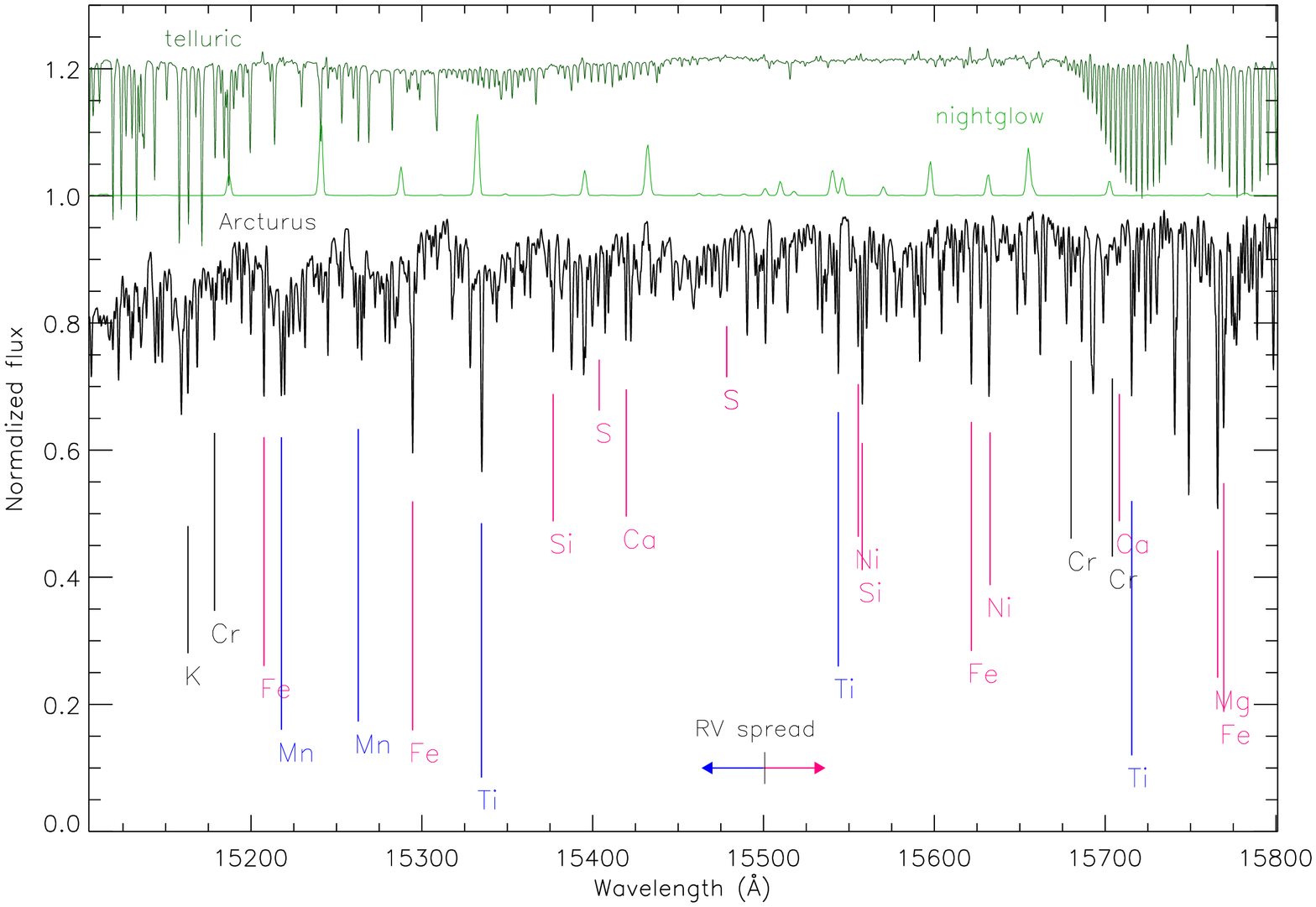}
\includegraphics[width=50mm,angle=90,height=5.7cm]{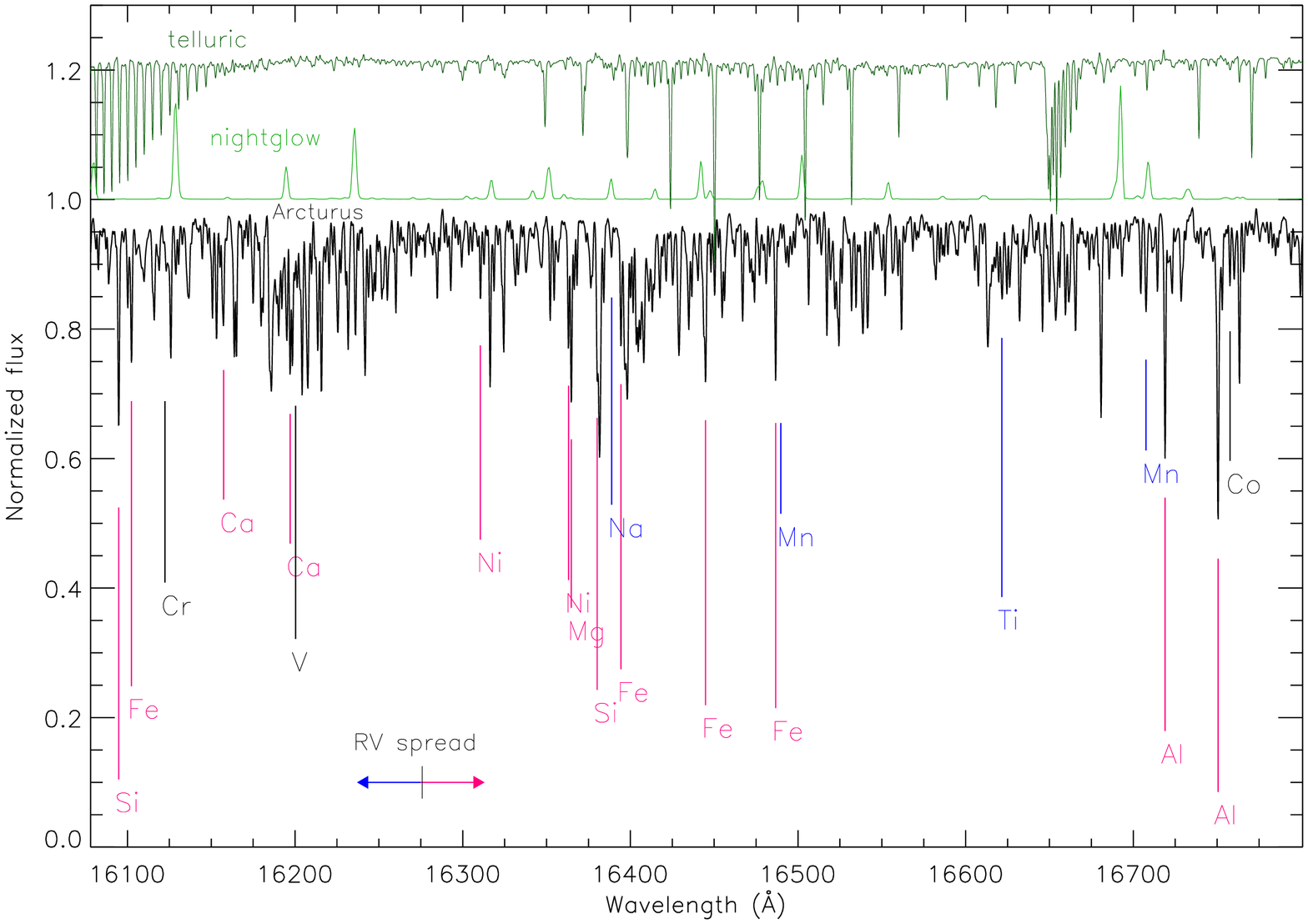}
\caption{Spectral ranges for APOGEE. 
The spectrum of Arcturus from the atlas
by Hinkle, Wallace \& Livingston (1995), smeared to a resolution of $R\simeq$ 20,000, is shown in black. Some of the strongest atomic lines are identified, color-coded according to their relevance for studying chemical evolution (red highest, blue high, black moderate). 
Numerous molecular bands (CO, CN, CH) also fall in this spectral window.  
The telluric absorption lines
and the nightglow spectrum, which have already been corrected for in the stellar
spectrum, are shown in green, vertically displaced and scaled. }
\label{label1}
\end{figure}

An $H$-band survey balances the need to efficiently punch through dust,
avoids a large thermal background, takes advantage of the peak of 
the red giant star spectral energy distribution, operates effectively during the more abundant and available 2.5-m bright time, and accesses a part of the spectral energy distribution that contains lines for multiple interesting elements.
Fig. 
\ref{label1} illustrates with the spectrum of Arcturus (K1.5 III) some of the lines available over the tentative 
spectral window for APOGEE. The figure also shows which regions are most
affected by telluric absorption/emission; these have already been removed
in the displayed stellar spectrum.

The goal of precision velocities and abundances determines that the instrument resolution be of
 order $R=$ 20,000. This resolution will enable 
an accurate location of the continuum, resolve crowded line packing, make possible efficient removal of 
telluric absorption/emission, and permit the derivation of abundances for key desired elements 
with $<0.1$ dex precision for stars as metal-poor as [Fe/H]$=-2.0$, 
provided a signal-to-noise ratio ($S/N$) of 100 per pixel.

\begin{table}
 \centering
\caption{Foreseeing performance of the APOGEE spectrograph}
\label{table}
\begin{tabular}{cc}\hline
Resolving power  & $\sim$20,000 \\ 
Nominal spectral coverage & 1.52 -- 1.58, 1.62--1.68 $\mu$m \\
Number of Fibers & 300 \\
Exposure time & 3 hr \\
$S/N$ & $>100$ \\
Limiting magnitude & $H=13$ \\
Detector & 
\begin{tabular}{c}
two 2048$^2$ \\
HgCdTe Raytheon chips
\end{tabular} \\
\hline
\end{tabular}
\end{table}

\begin{figure}
\includegraphics[width=150mm,angle=0,height=8.cm]{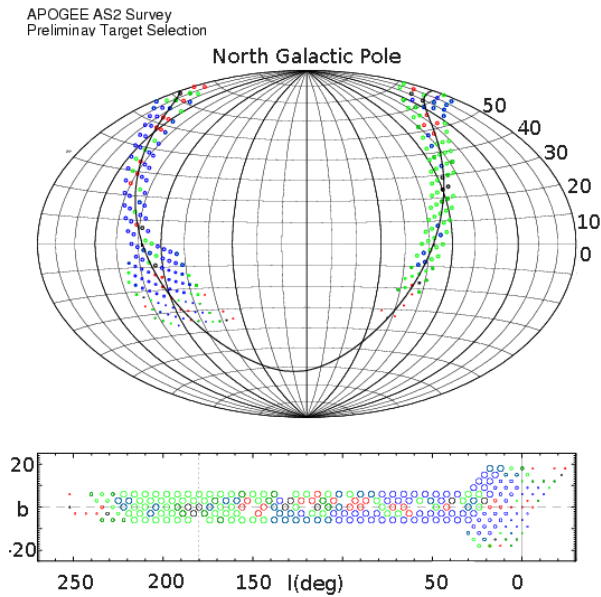}
\caption{One possible targeting strategy for low latitude APOGEE fields. Fields at the lowest declination
(shown by the smaller circles)
will be smaller because of differential refraction limitations, but there are plenty of available giant star targets
even in these reduced field areas. The colors indicate the target density in the fields:
red ($<$150), black (150--300), green (300--900), and blue ($>$900).}
\label{label2}
\end{figure}

 Typical target densities for potential APOGEE targets (e.g., red, evolved stars to $H\sim13$) within the SDSS telescope field-of-view are in the thousands towards the Galactic center, and 300-900 at the anticenter (see Fig. \ref{label2}). 
Survey efficiency pushes toward the maximal possible number of fibers, but maximal packing of spectra across 2048 expected detector pixels limits the number of fibers to about 300. 
This expected practical limit means that generally every fiber can be filled with a primary APOGEE target for low latitude pointings, and makes it practical to observe 100,000 stars in three years of bright time.  While $H\sim13$
is the nominal expected limit at $S/N=$100 for what will typically be three hour exposures, deeper probes
will also be possible for longer integration times, enabling access to the more distant halo as well as the most
deeply dust-buried disk stars.

The preliminary instrument design for APOGEE considers two Raytheon 
$2048\times2048$ HgCdTe detectors and
a VPH grating, sampling the 1.52--1.58$\mu$m and 1.62--1.68$\mu$m windows with a resolving
power $R\sim$ 20,000. A summary of the nominal instrument specifications
 is provided in Table \ref{table}. The entire instrument will operate at cryogenic temperatures
 to reduce the thermal background. Given the expected mass and size, it will be 
mounted on a bench with light fed from the 2.5m telescope by $H-$band optimized dry fibers. A bench-mounted spectrograph has the additional advantage of
improved stability, and more reliable calibrations (flatfields or
Th-Ar spectra).  Precise wavelength calibration will generally be afforded by the prominent airglow lines,
while telluric absorption will be mapped by hot stars included in each integration.

\section{Expected results}

Automated pipelines for data reduction will be used to process in 
a uniform fashion all the APOGEE spectra. The survey will generate
an estimated 15 Tb of raw data, and the processing and distribution
strategies will 
rely on the expertise accumulated by the SDSS  over the
last decade. 

The next step, the determination of radial velocities and chemical abundances,
will necessarily be performed by automated pipelines as well. Spectral synthesis,
 as opposed to integrated equivalent widths, will be used,  given the frequent
 overlapping of neighboring transitions. The ongoing
SEGUE survey (Sloan Extension for Galactic Understanding and Exploration;
Yanny et al. 2008, in preparation),
part of SDSS-II, precedes us, but the challenge for APOGEE will be to extend
its high-throughput data reduction pipeline to high-resolution 
spectroscopy.
A carefully devised calibration plan, including overlapping
fields and coordination with SEGUE, is also now under development.  

We anticipate public releases of APOGEE data including: (1) fully calibrated, 1-D spectra of each 
targeted star, (2) radial velocities precise to better than 1 km s$^{-1}$, (3) atmospheric
parameters ($\log g$, $T_{\rm eff}$, [Fe/H]), 
and (4) abundances 
for numerous elemental species.
The resulting catalog will be used to produce:
\begin{itemize}
\item A 3D map of abundances across the Galactic disk, bar, bulge and halo, 
	probing for correlations between chemistry and kinematics.
\item Constraints on the initial mass function and star formation rate of the bulge and
	the disk as a function of radius derived from, e.g., [$\alpha$/Fe] abundance trends.
\item A firm characterization of the bar of the Milky Way.
\item An accurate measurement of the Galactic rotation curve, which can be used to
constrain the dark matter density profile.
\end{itemize}

APOGEE will be diving into uncharted territory, and therefore unexpected
findings are to be expected, such as serendipitous discoveries of
peculiar, or extreme, stars and rare chemodynamical stellar populations.
The current road map calls for a very fast instrument development phase
 to be on the sky by 2011. 






\begin{thebibliography}{}
  \bibitem{} Allende Prieto, 
C. et al.: 2006, ApJ 636, 804 
  \bibitem{} Bensby, T., Zenn, A.~R., Oey, M.~S., Feltzing, S.: 2007, ApJ 663, L13 
  \bibitem{} Cunha, K., Smith, V.~V: 2006, ApJ 651, 491 
  \bibitem{} Fuhrmann, K.: 2008, MNRAS 384, 173 
  \bibitem{} Hinkle, K., Wallace, L., Livingston, W.: 1995, PASP 107, 1042 
  \bibitem{} Ivezi{\'c}, {\v Z}. et al.: 2008, ApJ 684, 287
  \bibitem{} L{\'o}pez-Corredoira et al.: 
 2007, AJ 133, 154 
  \bibitem{} Mel{\'e}ndez, J., et al.: 2008, A\&A 484, L21 
 \bibitem{} Reddy, B.~E., Lambert,  D.~L., Allende Prieto, C.: 2006, MNRAS 367, 1329 
  \bibitem{} Rich, R. M., Origlia, L.: 2005, ApJ 634, 1293
  \bibitem{} Skrutskie, M. F. et al.: 2006, AJ 131, 1163 
    \bibitem{} Wilkinson, M.I. et al.: 2005, MNRAS 359, 1306
  \bibitem{} York, D. et al., D.~G.: 2000, AJ 120, 1579 
  \bibitem{} Zwitter, T., et al.:  2008, AJ 136, 421 

  
\end{thebibliography}
\end{document}